\documentclass[a4paper,11pt]{article}
\usepackage{pos}
\usepackage{subcaption}
\usepackage{hyperref}
\usepackage{pgfpages}
\usepackage{fancybox,graphicx}
\usepackage{wasysym}

\usepackage{amssymb}
\usepackage{amsfonts}
\usepackage{amsmath}
\usepackage{relsize}
\usepackage{slashed}
\usepackage{wrapfig,rotating}
\usepackage[english]{babel}
\usepackage{tikz}
\usepackage{pgfpages}
\usepackage{fancybox,graphicx}
\usepackage{epstopdf}
\AppendGraphicsExtensions{.eps.gz}
\usepackage{times}
\usepackage[T1]{fontenc}
\def\leqsim{\mathbin{\;\raise1pt\hbox{$<$}\kern-8pt\lower3pt\hbox{$\sim$}\;}}
\def\geqsim{\mathbin{\;\raise1pt\hbox{$>$}\kern-8pt\lower3pt\hbox{$\sim$}\;}}
\def\p#1{\mbox{$ \mbox{\bf p}_1                                         $}}

\newcommand{\GeV}     {\mbox{$ {\mathrm{GeV}}                              $}}

\newcommand{\TeV}     {\mbox{$ {\mathrm{TeV}}                              $}}

\newcommand{\str}{\rm s \overline{s}}
\newcommand{\ba}{\begin{array}}
\newcommand{\ea}{\end{array}}
\newcommand{\bc}{\begin{center}}
\newcommand{\ec}{\end{center}}
\newcommand{\be}{\begin{eqnarray}}
\newcommand{\eeq}{\end{eqnarray}}
\newcommand{\bes}{\begin{eqnarray*}}
\newcommand{\ees}{\end{eqnarray*}}
\newcommand{\Kz}{\ifmmode {\rm K^0_s} \else ${\rm K^0_s} $ \fi}
\newcommand{\Zz}{\ifmmode {\rm Z^0} \else ${\rm Z^0 } $ \fi}
\newcommand{\xxbar}{\ifmmode {\rm x\bar{x}} \else ${\rm x\bar{x}} $ \fi}
\newcommand{\rphi}{\ifmmode {\rm R\phi} \else ${\rm R\phi} $ \fi}
\def    \missEt      {\ifmmode{/\mkern-11mu E_t}\else{${/\mkern-11mu E_t}$}\fi}
\def    \missE       {\ifmmode{/\mkern-11mu E}\else{${/\mkern-11mu E}$}\fi}
\def    \missp       {\ifmmode{/\mkern-11mu p}\else{${/\mkern-11mu p}$}\fi}
\def    \misspt      {\ifmmode{/\mkern-11mu p_t}\else{${/\mkern-11mu p_t}$}\fi}

\title % (optional, use  with long paper titles)
{Generating the full SM at linear colliders}
\manuallySeparateAuthors
\author*[\dag] % (optional, use  with lots of authors)
{Mikael Berggren}
\notes{\note{On behalf of the generator group 
(LCGG) of the Linear Collider Collaboration (LCC)}}
\affiliation% (optional, but mostly needed)
{
  DESY\\
  Notkestrasse 85\\
  D-22607 Hamburg\\
  Germany
}
\emailAdd{mikael.berggren@desy.de}
\abstract{Future linear e+e- colliders aim for extremely high precision measurements.
To achieve this, not only excellent detectors and well controlled machine conditions
are needed, but also the best possible estimate of backgrounds. To avoid that lacking
channels and too low statistics becomes a major source of systematic errors
in data-MC comparisons, all SM channels with the potential to yield at least a few
events under the full lifetime of the projects need to be generated, with statistics
largely exceeding that of the real data. Also machine conditions need to
be accurately taken into account. This includes beam-polarisation, interactions due
to the photons inevitably present in the highly focused beams, and coherent interactions
of whole bunches.
This endeavour has already been partly achieved in preparing design documents for both
the ILC and CLIC: Comprehensive samples of fully simulated and reconstructed events are
available for use.
In this contribution, we present how the generation of physics events at linear colliders
is categorised and organised, and the tools used. Also covered is how different aspects of
machine conditions, different sources of spurious interactions (such as
beam-induced backgrounds)  are treated and the tools involved for these aspects.

}
\FullConference{%
  40th International Conference on High Energy physics - ICHEP2020\\
  July 28 - August 6, 2020\\
  Prague, Czech Republic (virtual meeting)
}
\begin{document}
\maketitle
\tikzstyle{every picture}+=[remember picture]
\tikzstyle{na} = [baseline=-.5ex]
  \titlepage
\section{Linear colliders}
All proposed Linear colliders would collide polarised electrons with positrons.
Centre-of-mass energies would be ranging from 250 \GeV~ to 3 \TeV.
As the $e^+e^-$ initial state implies electroweak production, the background rates will be quite low at
such machines.
This has consequences for the detector design and optimisation. The detectors can feature close to $\sim 4\pi$ coverage,
and they do not need to be  radiation hard, so that the tracking system in front of calorimeters
can have a thickness as low as a few percent of a radiation-length.
In addition, the low rates means that the detectors needn't be triggered, so that {\it all}
produced events will be available to analysis.
Furthermore, at an   $e^+e^-$ machine, point-like objects are brought into  collision,
meaning that the initial state is fully  known.

Two options for linear colliders are currently under study, ILC and CLIC.
The  ILC \cite{Adolphsen:2013kya} has a defined 20 year running scenario, yielding
integrated luminosities of 2 and 4 ab$^{-1}$  at $E_{CMS}=$ 250 and 500 \GeV, respectively,
and would be up-gradable to 1 \TeV. An $E_{CMS}= M_Z$
option is also foreseeable. At the ILC, the positron beam would also be polarised.
To construct the ILC is currently under high-level political consideration in Japan.
Likewise, CLIC \cite{Aicheler:2012bya} has presented a 20 year staged running scenario,
yielding integrated luminosities of
5, 2.5 and 1 ab$^{-1}$ at  $E_{CMS}=$  3 \TeV , 1.5 \TeV,
and  380 \GeV, respectively.
CLIC is one possible future CERN project.
%  \subsection{MC requirements for linear colliders}

\begin{wrapfigure}{r}{0.35\columnwidth}
  \includegraphics [scale=0.25]{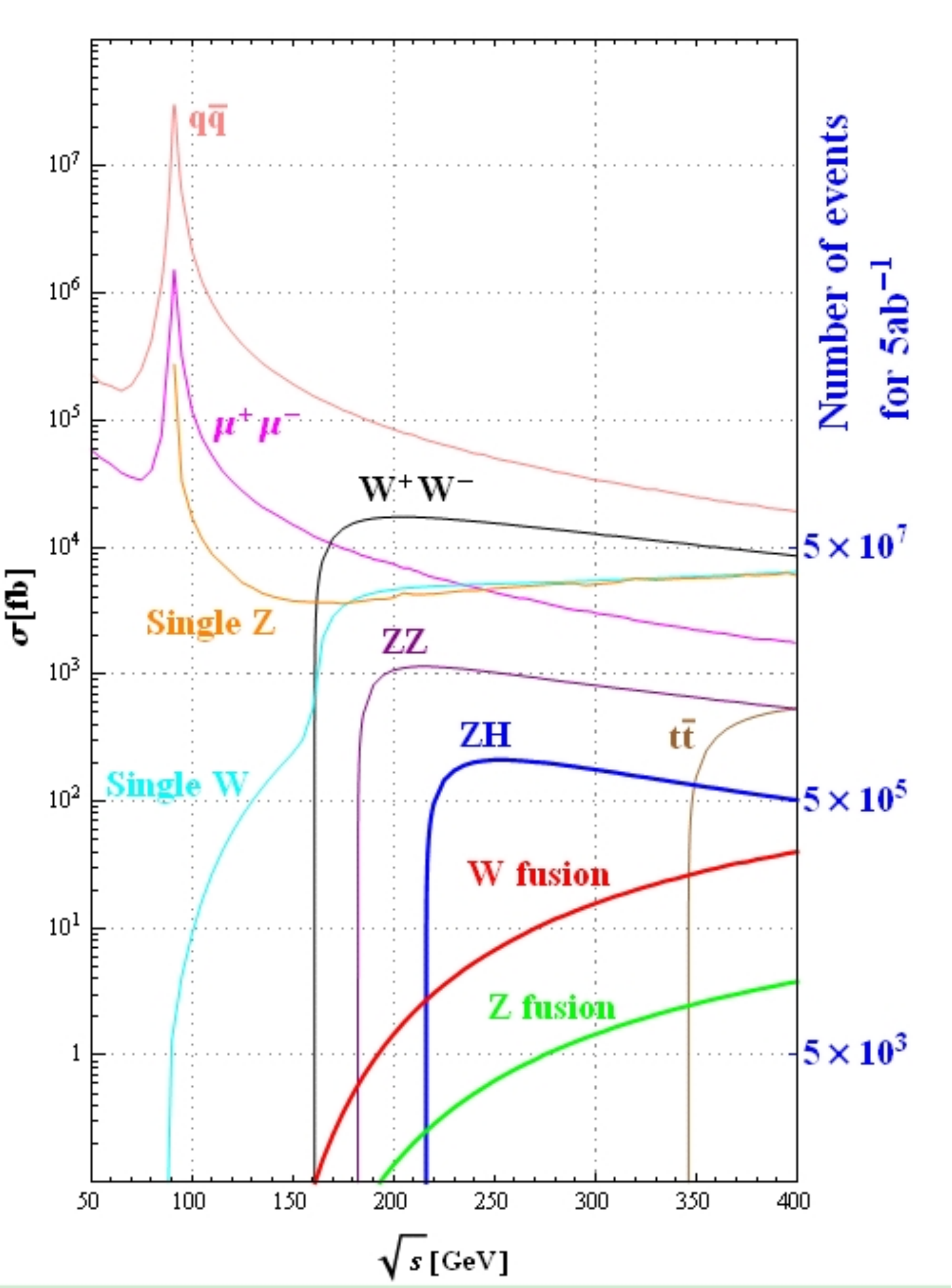}
   %    {{\tiny [from CepC, arXiv:1505.01008]}}
  \caption{Production cross-sections for various $e^+e^-$ processes. From \cite{Mo:2015mza}. \label{fig:xsects}}
 \end{wrapfigure}     
Linear colliders aim for extremely high precision measurements,
requiring  excellent detector performance, and well controlled machine conditions.
But it also requires the  {\it best possible estimate of backgrounds}.
A corollary to this is that  MC statistics or lacking
channels {\it must not} be a major source of systematic errors.
Therefore, all  SM channels yielding at least a few
events under the full lifetime of the projects need to be generated, with {\it statistics
largely exceeding that of the real data}.
In addition, machine conditions need to be accurately taken into account.
%This includes beam-polarisation, interactions due
%to the photons inevitably present in the highly focused beams, and coherent interactions
%of whole bunches.
Furthermore, at a linear collider, {\it all} events are interesting, and they are often
fully reconstructed.
In a sense, physics analysis at a linear collider might have more in common with
a B-factory than with the LHC.

The endeavour to achieve these requirements on event generation has been organised as a
common effort between the ILD and SiD detector concept groups at ILC \cite{Behnke:2013lya}
and CLICdp group at CLIC \cite{AlipourTehrani:2017gov}.
The work is done within the generator group, the \href{http://www.linearcollider.org/physics-detectors/working-group-softwarecomputing}{LCGG}, of the Linear Collider Collaboration (LCC).
%%xx The group 
%%xx has already partly achieved it's mission in preparing design documents for both
%%xx the ILC and CLIC: Comprehensive samples of fully simulated and reconstructed events are
%%xx available for use.
%  \end{itemize}
%\* {
%%x  {
%%x \begin{center}
%%x \begin{minipage}{0.6\linewidth}
%%x   \begin{center}
%%x     \begin{itemize}
%%x     \item  
%%x       Organisation and categorisation of generation of physics events.
%%x     \item  
%%x Treatment of
%%x machine conditions, different sources of spurious interactions
%%x %(such as
%%x %beam-induced backgrounds)
%%x %are treated
%%x     \item  the tools used. 
%%x    \end{itemize}
%%x   \end{center}
%%x \end{minipage}
%%x \end{center}
%%x }

\section{Generating the full SM\label{subsect:classification}}
%%xx At first, it might not seem too demanding to generate the full SM:
%%xx Once  an appropriate event generator has been found, it
%%xx should only be a question of acquiring sufficient CPU and disk-space resources,
%%xx resources that should be quite small compared to what the subsequent
%%xx detector simulation would require.
%%xx This is however not so.
To generate the full SM, there are many details to consider,
beyond the pure physics generation.
One must determine 
what is  colliding, since not only electrons and positrons are present in the beams, but also photons.
The incoming particles are not strictly mono-energetic, so the beam-spectrum must be
known and specified. The degree of polarisation of the beams must also be assigned,
and the distribution in space of the interaction point.
One must also consider what else happens during the beam-crossings,
namely
%the creation of e$^{+}$-e$^{-}$ pairs
beam-strahlung in the very strong
fields at the interaction point
%(``Beam-strahlung'').
%Finally,
%there are also
and 
parasitic $\gamma\gamma$ interactions in the
same bunch crossing as the physics events.
These sources of additional particles must be generated to be
able to correctly treat them in the subsequent detector simulation.
%\subsection{Process classification \label{subsect:classification}}
In addition to the specification of the beam-properties and determination
of spurious interactions, the physics channels themselves needs to be considered in detail,
when determining how to proceed with the generation.
At a linear collider, all events are interesting, and feature
a  huge spread in cross-sections, as illustrated in fig \ref{fig:xsects}.
At generation time, one cannot apriori know if 
a given physics study needs to consider a tiny cross-section process as
an important background,
or possibly a tiny fraction of a huge cross-section one.
The general characteristics of different sources of background
will be different in different cases.
Therefore,
processes should be grouped at generation time in
sufficiently well thought-through and well documented
way as to serve as many physics analyses as possible,
in the most convenient way.

\begin{figure}
  \begin{center}        
    \begin{subfigure}[b]{0.3\textwidth}
      \includegraphics [scale=0.1]{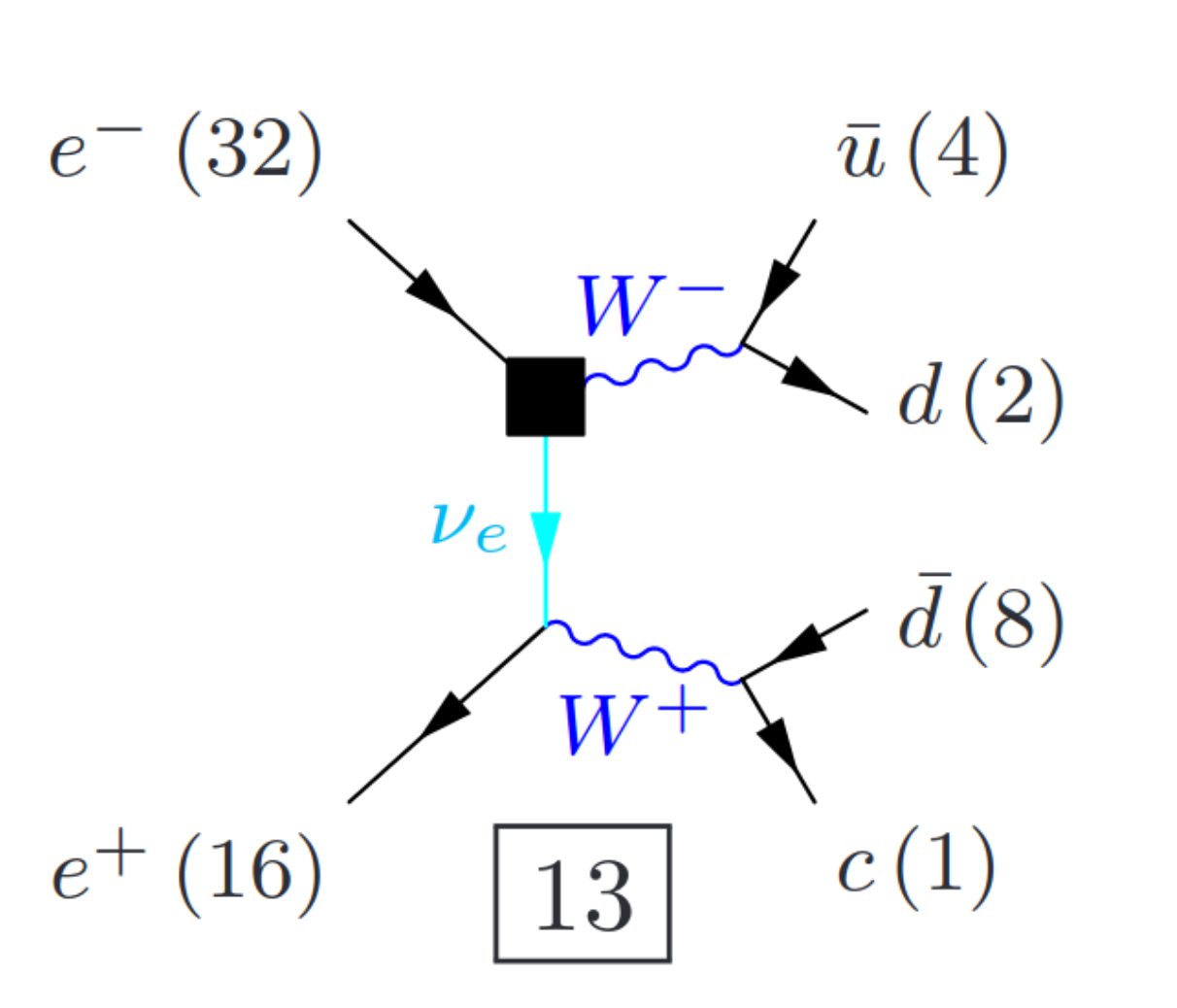}
       \caption{WW\_hadronic}
    \end{subfigure}  
    \begin{subfigure}[b]{0.3\textwidth}
    \includegraphics [scale=0.1]{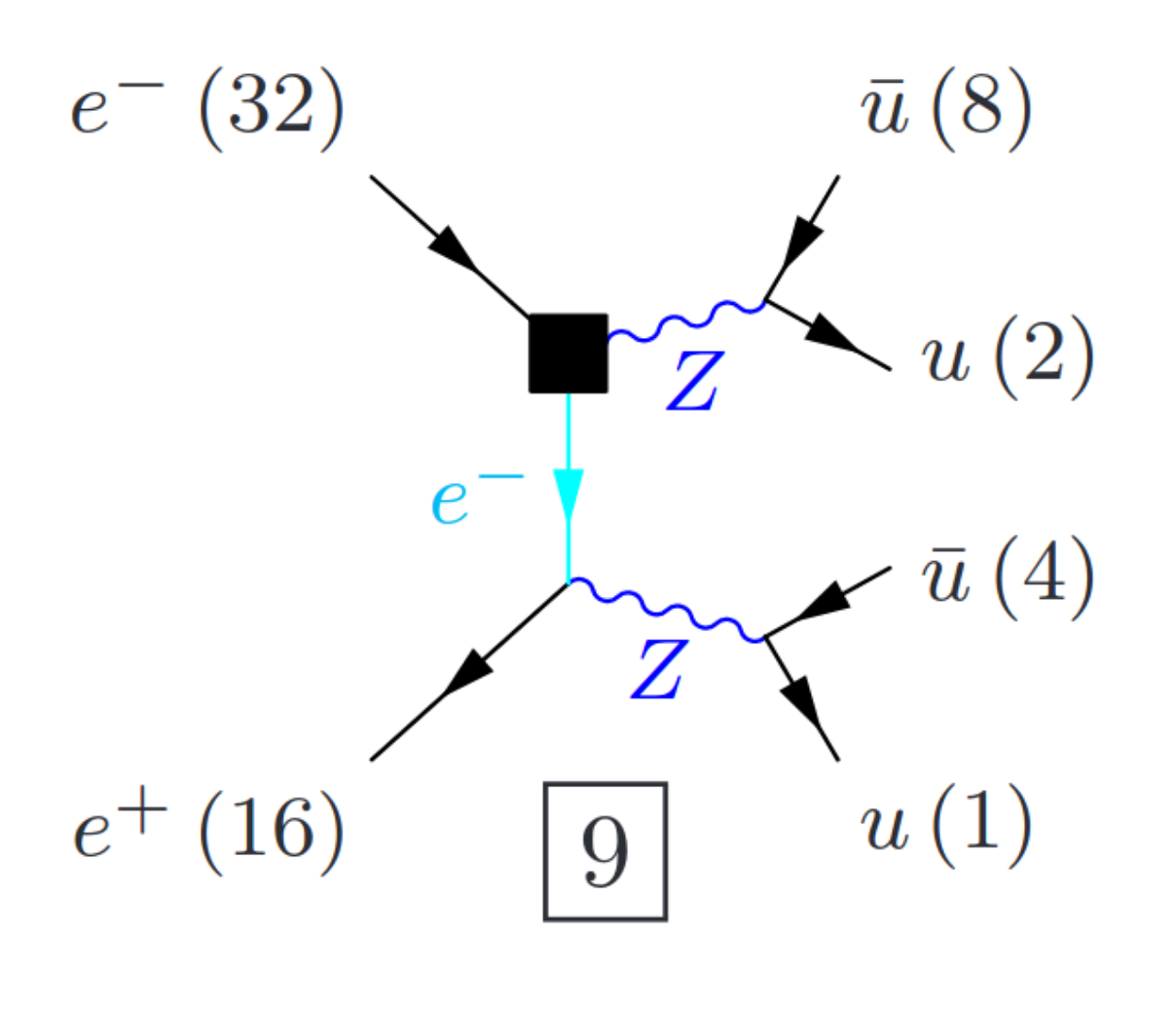}
        \caption{ZZ\_hadronic}
    \end{subfigure}  
    \begin{subfigure}[b]{0.3\textwidth}
   \includegraphics [scale=0.1]{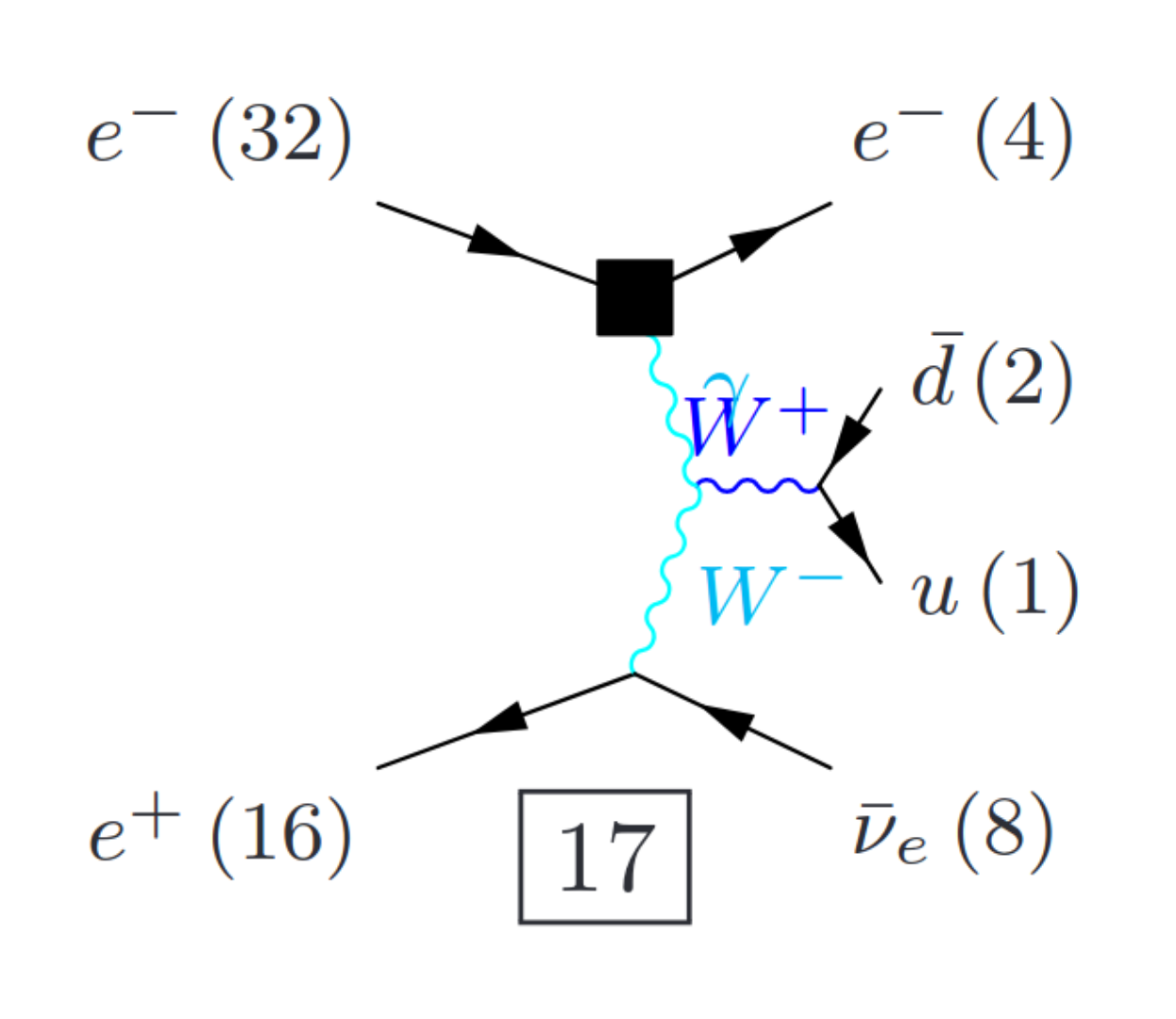}
        \caption{SingleW\_semileptonic}
    \end{subfigure}  
 \end{center}  
\vskip -0.5cm
  \caption{Examples of four-fermions diagrams with different flavour groupings.\label{fig:4fs}}
\end{figure}

 The process classification starts at defining the initial state, either e$^+$e$^-$, e$^{+(-)}\gamma$ or
 $\gamma\gamma$.
For electrons and positrons, the polarisation is specified, and for $\gamma$:s whether they are
real or virtual ones.
The    final state is then classified in several levels.
Firstly, the number of final fermions (1 to 8) is defined.
Then a  flavour-grouping is performed, determining if the final state can arise from
 intermediate W or Z bosons,
or if it is ambiguous.
Finally, fully  leptonic, fully hadronic, and semi-leptonic final states are separated.
In the ``Z-leptonic'' case , final states with neutrinos were also separated out.
Figure \ref{fig:4fs} illustrates a few examples of four-fermion flavour groupings.
%%xx \begin{figure}
%%xx   \begin{center}        
%%xx     \includegraphics [scale=0.12]{plots/whizard-channels_z_l0_onediag.png}
%%xx     \includegraphics [scale=0.1]{plots/ea_3f.png}
%%xx     \includegraphics [scale=0.08]{plots/aa_mumu.png}
%%xx   \end{center}  
%%xx   \caption{Different in-states}
%%xx \end{figure}
%%xx In addition, there are some    special considerations:
%%xx In  e.g. events with four fermions and  $\Sigma |L_e|$=2  in the final state,
%%xx t-channel production of single W or single Z dominates, yielding a quite different
%%xx event topology compared with events without t-channel contribution.
%%xx Double-counting must be  avoided, an example being  $\eeto  e^+e^- \gamma^* \gamma^* \rightarrow  e^+e^- f\bar{f}$
%%xx compared to $\eeto e^+e^- f\bar{f}$: these have the same final state, but are
%%xx generated differently.
%%xx In this case, it must be assured that the generator-level kinematic cuts cleanly separate the two
%%xx cases.
  \subsection{Main generator: Whizard}
{\tt Whizard} \cite{Kilian:2007gr} is the generator of choice for $e^+e^-$.
It features a full tree-level matrix-element evaluation.
It treats polarised beams, and contains a
full treatment of helicity densities.
The code traces the  colour flow in full, and passes this from the hard interaction generation
to the  code used to develop the subsequent parton-shower. 
Using it's {\tt Circe2} component,
% https://whizard.hepforge.org/circe2.pdf
{\tt Whizard} can handle arbitrary beam-spectra,
and,  using {\tt Tauola} \cite{Jadach:1990mz}, decays of polarised $\tau$:s
are correctly treated.
Finally, {\tt Whizard} can generate $2 \rightarrow 8$ processes.
%%xx which is  more important than NLO for  $e^+e^-$.
%%xx One can note that already for $e \gamma \rightarrow $ 5f, there are more than 500
%%xx diagrams contribution at tree-level.
The subsequent parton-shower and hadronisation is done by
other codes, typically {\tt Pythia6.4} \cite{Sjostrand:2006za}.
LCGG has tuned hadronisation using input from OPAL at
LEPII \cite{Boehrer:1996pr}.
%{\tiny [Phys.Rept. 291 (1997) 107-217, D. Ward, private communication.]}.

The process-definition given in the {\tt Whizard} steering file
(known as the {\it sindarin}
%% ;) \cite{Tolkien:silmarillion}
script, which is a component of the {\tt Whizard} package)
is also the driver for the scripts that organises the production along the lines
explained above.
Sindarin contains powerful grouping and aliasing capabilities, which are
exploited to assure that
no processes are over-looked.

\subsection{Generating beam properties}
The electron and positron beam-spectra are determined both by the incoming beam-spread,
and also by beam-beam interactions, which are due to the need to very strongly focus the
beams at linear colliders.
In addition to electrons and positrons, the beams also contain  photons, and
one must determine  how many they are, and whether they are  virtual or real.

\begin{figure}
  \begin{center}
    \begin{subfigure}[t]{0.4\textwidth}
    \includegraphics [scale=0.3]{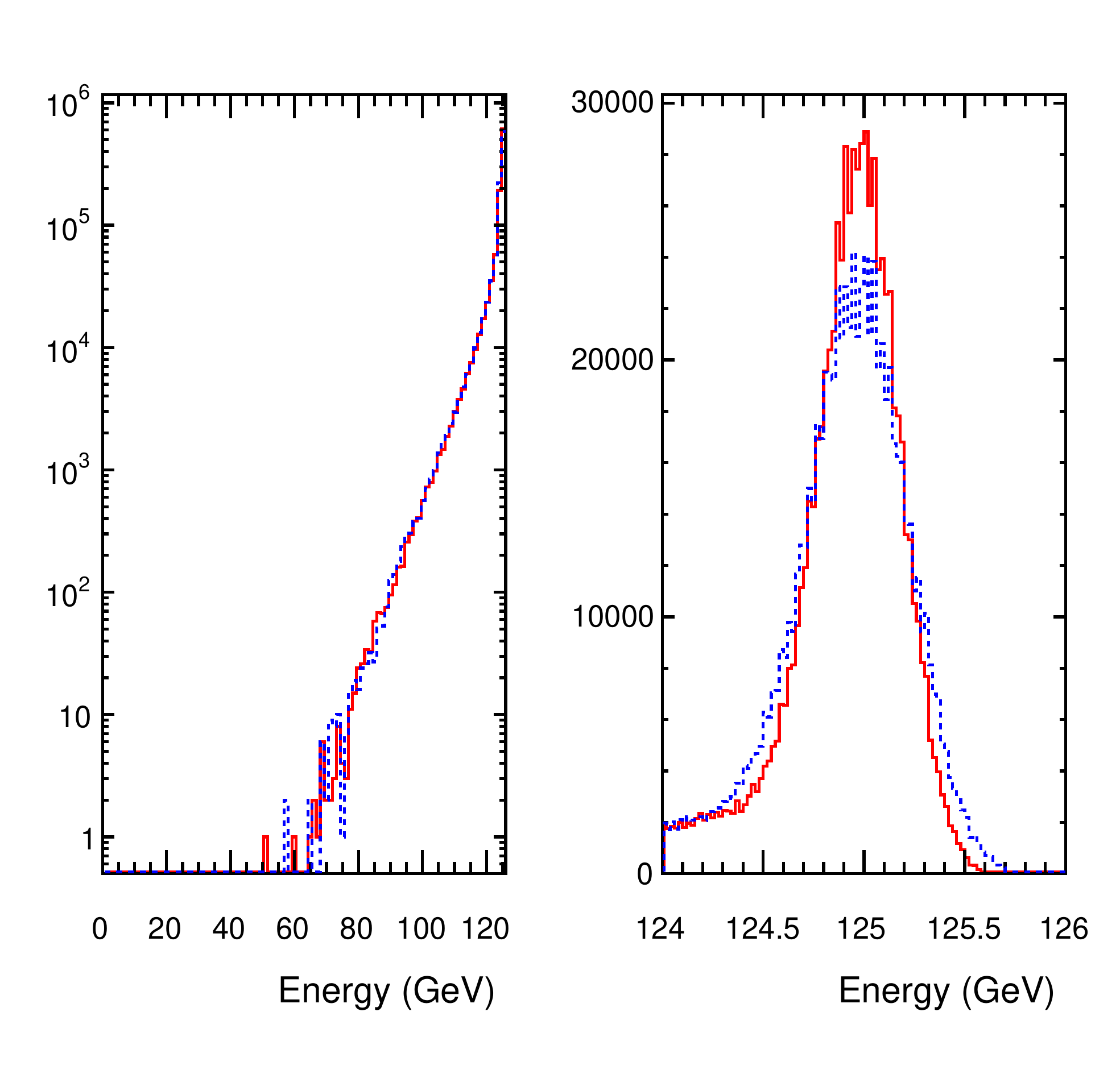}
     \caption{}
   \end{subfigure}
     \begin{subfigure}[t]{0.4\textwidth}
   \raisebox{0.55cm}{\includegraphics [scale=0.25]{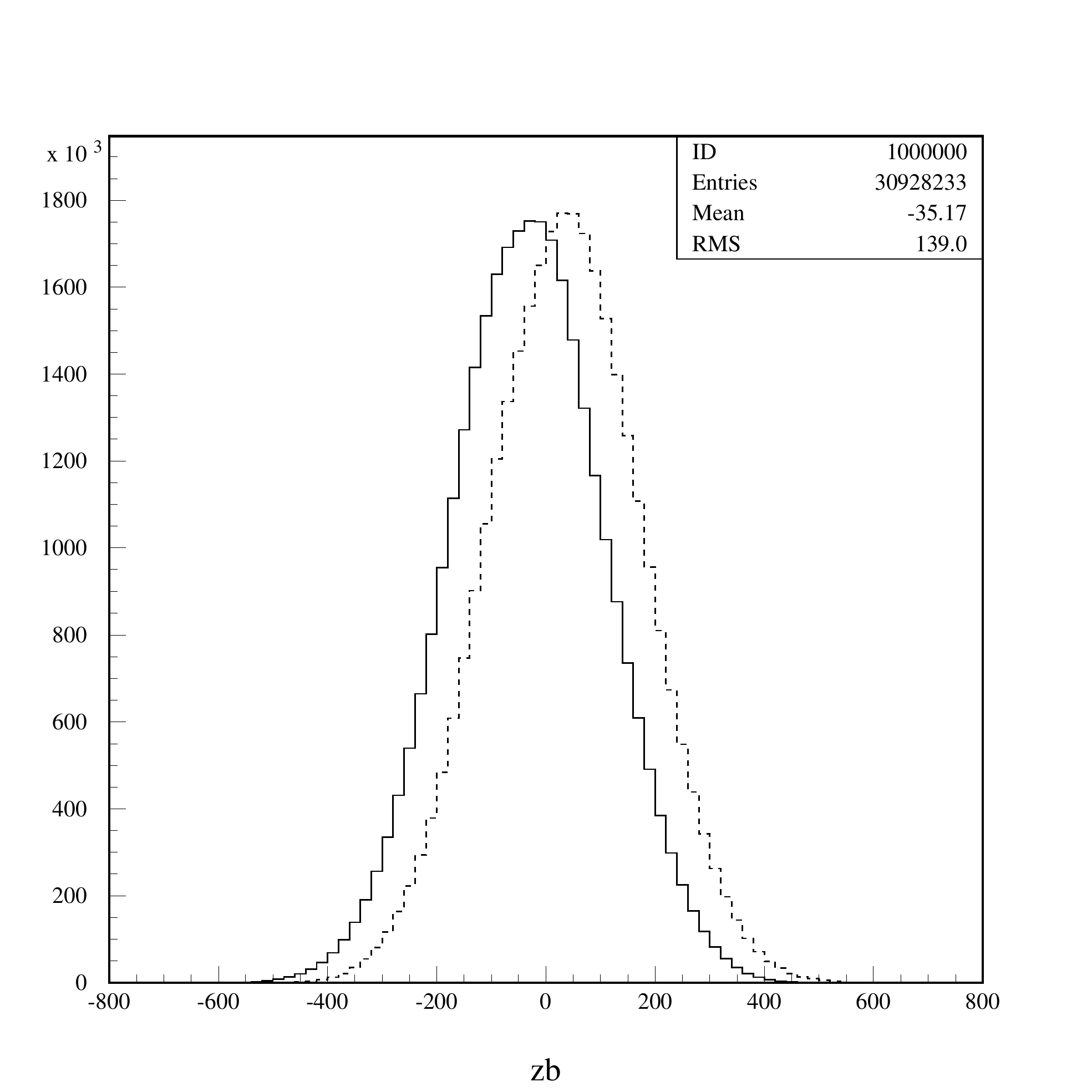}}
    \caption{}
  \end{subfigure}
  \end{center}  
\vskip -0.5cm
  \caption{Beam-characteristics:  (a) $e^-$ (dashed) and
    $e^+$ (solid) beam-spectra. (b) Position of interaction point for
    $e^+\gamma$ (solid) and $e^-\gamma$ (dash) events, in $\mu$m.\label{fig:beams}}
\end{figure}
The incoming beam-spread is due to the workings of the damping-rings and, for the electron beam, the ondulator (used for producing
the polarised photons needed to create the polarised positrons).
For these properties, external input from the machine-scientists are utilised,
%For the beam-beam interaction,  simulation input is needed.
while the interaction region is simulated using  {\tt GuineaPig} \cite{Schulte:1999tx}.
This program supplies the  beam-spectrum for electrons and positrons individually,
the amount and spectrum of real photons,
and the spacial distribution of the interaction point. Figure \ref{fig:beams} shows the
results of {\tt GuineaPig} for ILC-250.

In addition, there are spurious interactions of two types:
The  {\it pair-background}, which arises from pair-creation of photons in the beam by the strong fields.
The {\tt GuineaPig} code can also be used to generate the full activity during a beam-crossing (a {``BX''}).
%Typically, hundreds of thousands of such pairs are produced in each crossing.
The second type of  spurious interactions are {\it low-p$_\perp$ hadrons},
i.e. $\gamma^{(*)}\gamma^{(*)}$ interaction with small $M_{\gamma\gamma}$
and multiplicity, amounting to $\mathcal{O}(1)$ event in each BX.
%These are soft-QCD processes, which the ME procedure of {\tt Whizard} cannot produce.
These are either generated by {\tt Pythia},
%on the other hand, gives a good description down to $M_{\gamma\gamma} \sim$ 2 GeV.
%For  $M_{\gamma\gamma}$ below this point,
or a custom generator developed by LCGG
relying on fitting to available data.
%was used 
For  both these types of spurious interactions
%the same procedure
%was used, when integrating them into the detector simulation.
a pool of events were pre-generated, from which events were
picked at random and overlaid on the main event during detector simulation.
In the case of  the  pair-background, the pool of events where
obtained by simulating  $\sim$ {10$^5$ BXes}, and using 
the    fast detector simulation SGV \cite{Berggren:2012ar}
%\href{https://www.desy.de/~berggren/sgv_ug/sgv_ug.html}{\underline{SGV}}
to select those tracks actually reaching any element of the tracking system ($\sim$10/BX).
A large fraction of the rest of the pairs
%- those not disappearing into the beam-pipes -
would hit the very-forward calorimeter system (the BeamCal),
and were used to build a map of background on the system,
%This map was
used in the full detector simulation to simulate the
BeamCal.
%%xx In the case of the low-p$_\perp$ hadrons, the events generated were added to the pool in full.
%%xx Differently from the  case of  the  pair-background, the full simulation would pick a Poisson distributed random amount of such
%%xx events to overlay each physics event.

  \section{Setup, integration, event generation, and documentation}

  The generation is preformed in a number of steps,
with verification done between them. The initial few steps are done interactively.
First, the     {\tt Whizard} process definition is parsed to build a directory-tree structure, with one unique directory
per process. This is to avoid possible race-conditions later, in the actual generation step.
%% Then
The process-specific code is generated and compiled,
%This is done for each class of same initial state and number
%of final state fermions; it is not needed to do this separately for different flavour combination.
%This is also done interactively.
%% Subsequently, the directory tree is traversed to do an  interactive ``pre-integration'' of all channels,
%% to flag zero cross-section ones.
%Subsequently,
%and
then the tree is traversed to do a ``pre-integration'' of all channels,
to flag zero cross-section ones.

The full integration of all  (non-zero cross-section) channels is then submitted to the
local batch-farm,
%under Condor.
with the goal that the calculated cross-section should be accurate to 0.1 \%.
For all channels with  $\le$ 5 fermions, this was found to be an over-night job.
The last step before the full generation was a ``pilot generation' of all channels, with 1000 events/channel.
This was useful to  evaluate the CPU time and storage needed for the full generation.
%% as well as verifying that there
%% were no technical problems needing attention in any channel.
A number of channels were identified
for which either the precision or
efficiency of the generation was problematic.
These were channels with low cross-sections and complicated final states,
and it was decided to defer their generation until after the full simulation
of the bulk of the samples was completed.
%
%      At this point, the system is ready to generate the full SM.
%  \subsection{Generation}

\begin{figure}
  \begin{center}
    \begin{subfigure}[t]{0.4\textwidth}
    \includegraphics [scale=0.18]{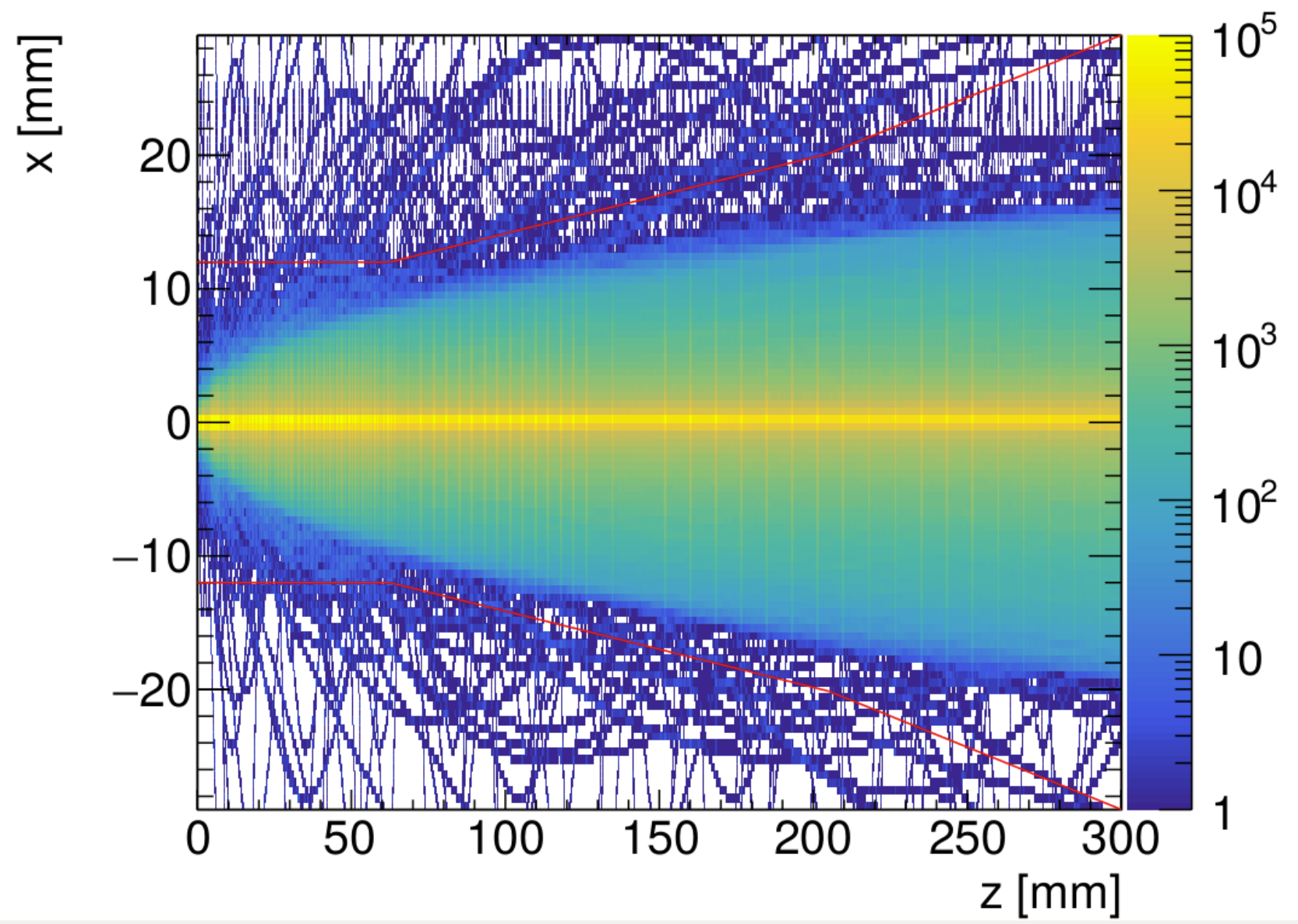}
     \caption{Pair-background in one BX.}
   \end{subfigure}
    \begin{subfigure}[t]{0.4\textwidth}
    \includegraphics [scale=0.38]{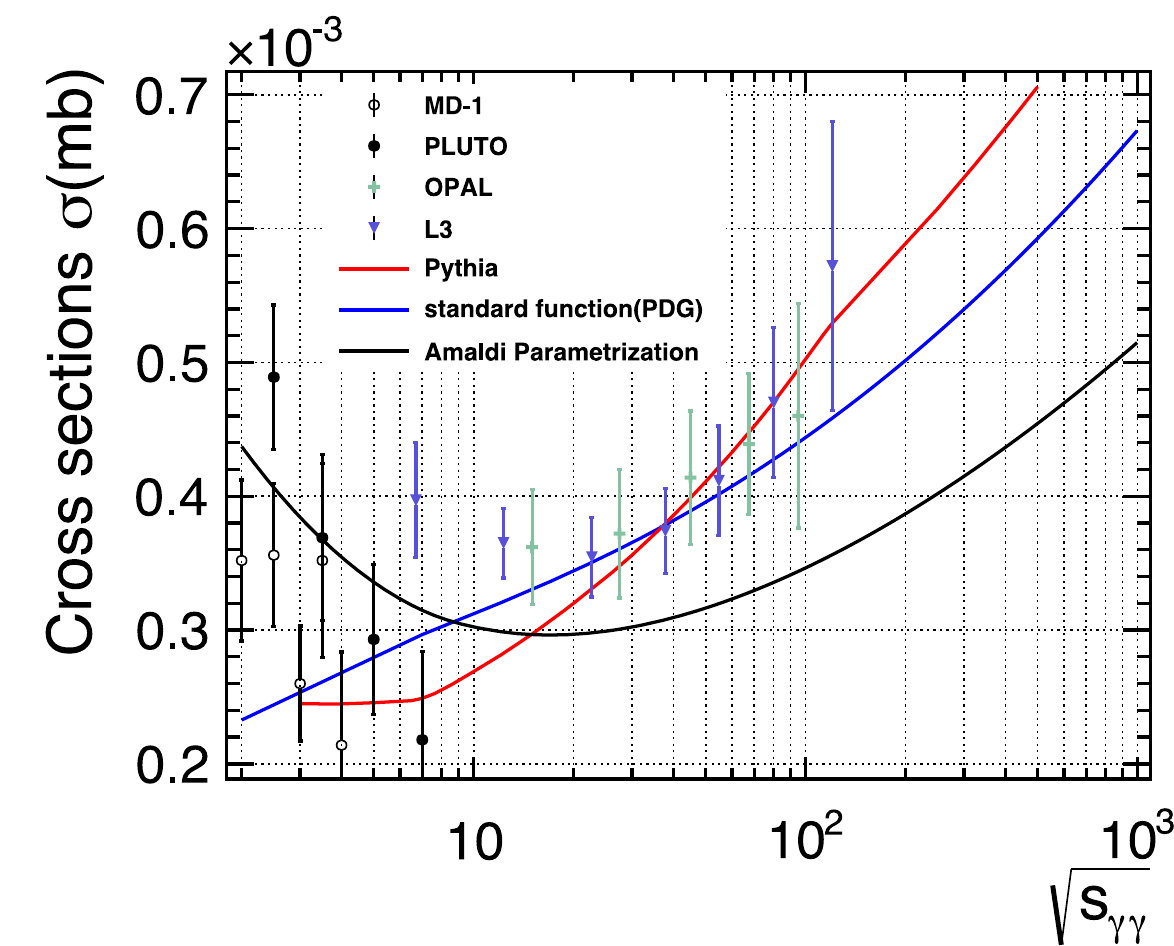}
     \caption{Cross-section of $\gamma\gamma \rightarrow hadrons$.}
   \end{subfigure}
  \end{center}  
\vskip -0.5cm
  \caption{Sources of spurious particles. In (a), the red line is the beam-pipe.\label{fig:overlay}}
\end{figure}
The full generation was also done on a batch-farm.
%In most cases, one process constituted one generation job.
%However, some processes alone represent $\sim$ billion events. These must be split in several jobs.
%During the jobs, a daemon was run to identify completed outputs.
The production of event files - in LCIO \cite{Aplin:2012kj} format -
was supervised by a daemon, which was responsible to upload each completed
file to the grid as soon as possible.
%As soon as a file - in LCIO \cite{Aplin:2012kj} format - was found to
%be complete, it was uploaded to
%the grid.
Upon successful upload, the local file was deleted, avoiding
disk-space issues on the batch-cluster.
%% the risk
%% that the jobs would run out of disk-space on the batch-cluster.
%% If the upload was successful, the local file was deleted, avoiding the risk
%% that the jobs would run out of disk-space on the batch-cluster.
%At the end of each job, the events (in LCIO \cite{Aplin:2012kj} format),
%    (in \href{https://github.com/iLCSoft/LCIO}{\underline{{LCIO}}} format),
%metadata, and input+logfile tarballs are uploaded to the grid.
When {\it all} jobs of a channel were completed, a summary metadata file was created
and uploaded to the grid, together with input and log-file tarballs.
The existence of the metadata file triggered the
full simulation system
%% simulation and reconstruction system 
under {\tt DIRAC} \cite{Tsaregorodtsev:2008zz}
%\href{http://diracgrid.org/}{\underline{{DIRAC}}}.
to start processing the channel.

    A number of precautions were taken to balance the size of the output files,
both in physical size, and the number of events they contained: a demand from the down-stream
detector simulation was that all files of any given process should contain the same number of events,
and that single files should not (much) exceed 500 MB in size. As channels were split both in several files,
but also in some cases in several jobs, care should be taken with event- and file-numbering: no gaps in the event-numbering
sequence, nor in the file sequence numbers. 

%\subsection{Generation status}
%For $\le$ 5f (w/o virtual $\gamma$ induced ones):
At the time of writing (December 2020),
104 of the total of 212  channels with $\le$ 5 fermions have been generated (478 jobs),
corresponding to  2.7 billion events -  5.4 TB in 15788 files.
The total CPU time was 7233 hours, which was obtained in 10 days on the batch farm.
%The remaining channels are those that either
The remaining deferred channels
%which are those with the low cross-sections and complex final states
are expected to
contain 0.5 billion events.
%% , and will be produced shortly.
%%%% xxxx  xxx   yyyyy yyyy   zzz zzz   dddd dddd   aaaa aaaaaa   ccccc cccc
%%%
%%% xxxx  xxx   yyyyy yyyy   zzz zzz   dddd dddd   aaaa aaaaaa   ccccc cccc
%%xx As of today, {\it104} channels are done, producing  {\it 2.7}~billion events in {\it 15788} LCIO files
%%xx occupying {\it 5.4~TB}.
%%xx This used {\it 7233 CPU hours}, obtained in $\sim$ 10 days.
%%xx The remaining channels are
%%xx  only $\mathcal{O}(10000)$ events. 
%%xx   In addition, there are {\it 96} channels with virtual $\gamma$:s to come, $\sim${\it 0.5}~billion events.
%%xx  In most cases: one channel = one generation job, but in some processes alone represent {\it $\sim$ billion events}:
%%xx     split in several jobs. In total, {\it 478} jobs have been completed.
%%xx  At the end of each {\it job}, the {\it events}
%%xx     (in LCIO format),
%%xx     metadata, and input+log-file tarballs are {\it uploaded to
%%xx       the grid}.
%%xx     At the end of each {\it  channel}, {\it summary metadata} of all jobs of the channel
%%xx     are uploaded to the grid.
%%xx     This triggers the simulation and reconstruction system 
%%xx     under DIRAC to do it's thing.
%%xx     This  documentation is also available on {\it the Web}.
  
%\subsection{Documentation}
The full documentation of each channel was created by  generation job, driven by the contents of
the  process- definition Sindarin script
and common conditions. This information can be found in
the header of each event (Process-id, beam-polarisation, and cross-section),
and, in full, in the generator  meta-data files.
%The latter condenses job-specific information from  the Whizard logs,
%and  contains the process (both in short-hand, and in a more human-readable form), the cross section, the beam polarisation,
%file-names, number of events and file, the corresponding
%integrated luminosity, beam-spectrum used, and a number of technical information on version numbers.
Apart from being uploaded to the grid, this information is also available in browasble format
on the \href{https://ilcsoft.desy.de/dbd/generated/}{Web}.
In addition to the condensed metadata, all
steering-files, log-files, pdf:s showing the diagrams contributing, and the integration phase-space grids,
are available on the Web, 
and in tar files, which are stored on the grid in a parallel directory to the generated files.
This collection of information is sufficient to re-do the generation, if e.g. more events are
requested in some channel, or if it is found that detailed debugging would be needed.

\section{Conclusions}
  We have explained that the precision-goal of  future $e^+e^-$ colliders are such that
  it is {\it not permissible} that MC statistics or coverage could constitute an {\it major systematic uncertainty}.
  In this spirit, we showed how the generation of the full SM can be achieved.
  It consists of bringing a large number of different codes together, namely:
  {\tt Whizard}, {\tt Pythia}, and {\tt Tauola} for physics generation,
  {\tt GuineaPig}, and  {\tt Circe2} for beam-properties, and
  {\tt SGV+GuineaPig} and the Peskin/Barklow generator for spurious interactions.
In addition, input from machine-physics and data from LEPII was used.
  This full data is organised and documented in a physics-oriented fashion, for the benefit of
    the end-user.
    The
    %entire
    system pivots around {\it one data-source}, the Whizard process definition file.

\end{document}